\documentclass[11pt,aps,preprint,nofootinbib,floatfix]{revtex4}

\usepackage{epsfig}
\usepackage{bbm}                                                  
\usepackage{bm}

\newcommand{\nc}{\newcommand}
\nc{\beq}{\begin{equation}}  \nc{\eeq}{\end{equation}}
\nc{\bea}{\begin{eqnarray}}  \nc{\eea}{\end{eqnarray}}
\nc{\baa}{\begin{array}}     \nc{\eaa}{\end{array}}
\nc{\bit}{\begin{itemize}}   \nc{\eit}{\end{itemize}}
\nc{\ben}{\begin{enumerate}} \nc{\een}{\end{enumerate}}
\nc{\bce}{\begin{center}}    \nc{\ece}{\end{center}}
\nc{\bpm}{\begin{pmatrix}}   \nc{\epm}{\end{pmatrix}}
\nc{\bvt}{\begin{verbatim}}  \nc{\evt}{\end{verbatim}}

\def\dm#1{{\rm diag}\left(#1\right)}
\def\up#1{^{\left( #1 \right) }}
\def\su#1{{SU(#1)}}
\def\ui{U(1)}
\def\vevof#1{\left\langle#1\right\rangle}
\def\vev{vacuum expectation value}
\def\leaderfill{\leaders\hbox to .3em{\hss.\hss}\hfill}
\def\mubf{{\bm\mu}}
\def\mati{{\mathbbm1}}
\def\acal{{\cal A}}
\def\bcal{{\cal B}}
\def\dcal{{\cal D}}
\def\lcal{{\cal L}}
\def\mcal{{\cal M}}
\def\ocal{{\cal O}}
\def\qcal{{\cal Q}}
\def\xx{{\bf x}}
\def\mBB{{\mathbbm M}}
\def\rBB{{\mathbbm R}}
\def\vBB{{\mathbbm V}}
\def\inv#1{\frac1{#1}}

\def\tev{\hbox{TeV}}
\def\lesim{\,{\raise-3pt\hbox{$\sim$}}\!\!\!\!\!{\raise2pt\hbox{$<$}}\,}
\def\half{\frac12}

\begin{document}

\preprint{IFT-26-05}

\preprint{UCRHEP-T400}

\title{Light excitations in 5-dimensional gauge theories}

\thanks{Presented at the
XXIX International Conference of Theoretical Physics Matter To The Deepest:\vspace{-.1in} \\ 
Recent Developments In Physics of Fundamental Interactions Ustron, 8-14 September 2005, Poland\vspace{.1in}}

\author{Bohdan GRZADKOWSKI}
\affiliation{Institute of Theoretical Physics,  Warsaw University Ho\.za 69, PL-00-681 Warsaw, Poland}
\email{ bohdan.grzadkowski@fuw.edu.pl}

\author{Jos\'e WUDKA}
\affiliation{Department of Physics, University of California, Riverside CA 92521-0413, USA}
\email{jose.wudka@ucr.edu}

\begin{abstract}
We consider general five-dimensional gauge theories compactified on 
on an orbifold $S_1/Z_2$ 
with all fields propagating in the bulk. We propose a generalized 
set of boundary conditions and derive the general features
of the low energy-spectrum. The results are illustrated with
two simple examples.
\end{abstract}

%\pacs{11.10.Kk, 11.30.Er, 11.30.Qc, 12.20.Ds}
  
\maketitle

\section{General features}

Theories in $ n> 4 $ dimensions are based on solutions
(assumed or exhibited) to the $n$-dimensional Einstein
equations that contain $ n-4$ compact dimensions whose
typical size we denote by $L$. 
These models can be conveniently divided into 
``large'' and ``small'' extra dimensional theories,
subdivided into models containing branes
and those that not. 

Large extra dimensional theories~\cite{Arkani-Hamed:1998rs}
assume $ L$ to be of sub-millimeter-size and 
that all fields but gravity are confined
to a 4-dimensional subspace (the ``brane''). In these models
the electroweak
scale $v$ is the only energy scale, and the Planck mass
is a derived quantity
equal to $ M_{\rm Pl} = v ( v L)^{(n-4)/2}$. However, $ L v \gg1$
is also required, which can
be maintained only through fine tuning. In addition there
is no inclusion of the brane-induced gravitational 
effects and, finally, there are complications when
implementing the confining mechanism.

The simplest model with small extra dimensions 
containing branes~\cite{Randall:1999ee}
is obtained from an explicit solution to the Einstein
equations with one or two branes,  assuming that the
main brane  contribution to the energy momentum tensor comes
from the brane cosmological constants. This model (and
its extensions) have the virtue of relating the Plank
and weak scales through a metric-induced exponential conformal
factor that naturally implements the hierarchy $ G_F M_{\rm pl}^2 \gg 1 $ 
when $ L \sim 1 \tev^{-1} $.
This, however, is achieved at a price: the brane and bulk cosmological
constants must be appropriately tuned
to achieve this effect. In addition the perturbative
expansion around the solutions obtained produces a zero
mode, indicating that the obtained configuration is
marginally stable.

Finally, the ``universal'' extra-dimensional models~\cite{Appelquist:2000nn}
also assume small extra dimensions 
($L\lesim 1\tev^{-1}$) but now without branes;
the compact directions are flat  and that all fields propagate
throughout the $n$-dimensional space. These models avoid
phenomenologically unacceptable deviations from low-energy
physics  because of the absence of vertices containing
a single heavy leg (a consequence of momentum 
conservation)~\cite{Appelquist:2000nn}. Such theories 
contain dimensional non-renormalizable couplings
(as all higher-dimensional theories) which imply the presence
of an energy scale $ \Lambda $ (the cut-off)
beyond which the theory cannot be applied (at least perturbatively).
Despite this such models have the virtue of containing scalars whose
masses do not suffer form $ O(\Lambda)$ corrections, these
being instead $ O(1/L)$~\cite{Hatanaka:1999sx},\cite{Hosotani:1988bm}.

In this talk we will consider a 5-dimensional universal model containing
only gauge-fields and fermions. We will describe a very  general
type of behavior for the fields under the symmetries of the
compact subspace, and derive some of the associated consequences,
concentrating on the possible light spectra present in such
models. These features are then illustrated with 2 examples
(we do not address the stability of the assumed space-time
 configuration, nor do we consider any gravitational effects).
The ultimate goal of these models is  to construct
a realistic theory without including fundamental 
(5-dimensional) scalars; as 
far as the authors know such model does not yet exist, still,
we hope to show that these theories are sufficiently interesting
to warrant further study.

\section{The Lagrangian}

The Lagrangian is assumed to have the form
\bea
\lcal = - \inv4 \sum_a \inv{g_a^2} \left( F_{MN}^a\right)^2
+ \bar\Psi \left( i \gamma^N D_N - M \right) \Psi ,
\eea
where all fermions have been lumped in a large multiplet
$ \Psi$, the covariant derivative equals
$ D_N = \partial_N + i g_5 A_N^a T_a$ where 
$ g_5\sim({\rm mass})^{-1/2}$ (which is the dimensional
coupling mentioned previously) and the $T_a$ generate the
(in general reducible) representation carried by the
fermions. The gauge coupling constants have been written as
$ g_a g_5 $ with $g_a$ dimensionless, and the gauge fields
were then re-scaled appropriately; the $g_a$
have the same value for all indices $a$ within the same factor
group. $M,~N,\ldots=(0,1,2,3,4)$ denote 5-dimensional space-time indices 
with the first four corresponding to the usual Minkowski space
(labeled by Greek letters $\mu,~\nu, \ldots$); the
last index corresponds to the compact direction and we use $ x^4=y$.

Considering the most general properties of this model
in a compact space it proves convenient to define a
fermionic multiplet $\chi$ by
\beq
\chi = \bpm{ \Psi \cr - \Psi^c }\epm ,
\eeq
where
$\Psi^c = C \bar\Psi^T,~C=\gamma_1 \gamma_3 $.
In terms of $\chi$ we find
\beq
\lcal = - \inv4 \sum_a \inv{g_a^2} \left( F_{MN}^a\right)^2
+ \half \bar\chi \left( i \gamma^N \dcal_N - \mcal \right) \chi ,
\label{eq:defs}\eeq
where
\beq
\dcal_N = \partial_N + i g_5 A_N^a \tau_a ;
\quad
\tau_a = \bpm{ T_a & 0 \cr 0 & -T_a^* }\epm ,
\quad
\mcal = \bpm{ M & 0 \cr 0 & -M^* }\epm .
\label{eq:defs2}\eeq
$\lcal$ is invariant under P and C discrete symmetries
defined by~\footnote{In terms of the usual Dirac matrices we chose
$ \gamma_N = (\gamma_0, \gamma_1,\gamma_2,\gamma_3,i\gamma_5)$.}
\bea
{\bf P:} &&
(x^0, \vec\xx,x^4) \to (x^0, {\bf-}\vec\xx, x^4); 
\quad \chi \to \gamma_0 \gamma_4 
\chi; \cr
{\bf C:} &&
\chi \to \chi^c = - i \sigma_2 \chi 
\eea

In writing the Lagrangian in terms of $\chi$ we must insure that
no new degrees of freedom are introduced; this is implemented
by the constraints
$\chi = i \sigma_2 \chi^c$,
$\sigma_{1,2}\tau_a \sigma_{1,2}= - \tau_a^*$, 
$[\sigma_3 , \tau_a ] =0 $,
which follow form the definitions~\footnote{In these
expressions the $\sigma_i $  have the
standard form except that the entries are
replaced by unit and zero (square) matrices of
size equal to the dimension of $ \Psi $.}.

\section{The 5-dimensional space time}

We consider a space of the form
$ \mBB \otimes (\rBB/\qcal)$ where $ \mBB$ denotes
the 4-dimensional Minkowski space-time and
$\qcal$ is a discrete group with two elements
($x^4=y$ denotes the coordinate of $\rBB$):
{\it(i)} Translation, $ y \to y+L $, where $L$ denotes
the size of the compact subspace; and {\it(ii)} reflection,
$ y \to -y $. Both of these act trivially 
on $ \mBB$~\cite{Quiros:2003gg}.

Under translations we assume that the fields 
transform according to~\cite{Grzadkowski:2005rz}
\bea
\Psi(y+L) &=& \Gamma \Psi(y) 
+ \Upsilon^* \Psi^c (y) \cr
A_N^a(y+L)  T^a &=&
A_N^a(y) \cdot \left\{
\begin{array}{lr}
U_1^\dagger\left( T_a \right) U_1   & (P1)\cr
U_2^\dagger\left(- T_a^*\right) U_2 & (P2)
\end{array} \right.
\eea
where $ \Gamma $ and $ \Upsilon$ are constant matrices
and $ U_i,~i=1,2$ constant gauge transformations. The
above expression is a generalization of the usual 
assumptions which correspond to 
choosing P1 and $ \Upsilon =0 $; the
possibility of having non-vanishing $ \Upsilon $ stems
from the charge symmetry of the original theory. It
is clear, however, that this matrix can relate only
components in $ \chi$ that correspond to non-complex
representations of the gauge group (else gauge invariance
would be compromised). The possibility of having the
transformation $P2$ for the gauge fields is suggested
by that of the fermions; in contrast with these, however,
no linear combination of $P1$ and $P2$ is allowed since it
does not leave the $F^2$ terms in the Lagrangian invariant.

The observation that one can add transformation rules
involving $ \Upsilon$ and/or $U_2$ is one of the main
point of this talk. The presence of these terms allows
for a much richer phenomenology in these theories and,
in particular, for wide variety of spectra in the low 
energy theory.

In terms of $ \chi$ the above expressions become
\beq
A_N^a(y+L)  = \vBB_{ab} A_N^b(y); \qquad
\chi(y+L) = \acal \chi (y) ; ~~ \acal = \bpm{\Gamma & -\Upsilon^* \cr \Upsilon & \Gamma^* }\epm
\label{eq:transl}
\eeq
where $ \vBB$ (whose sub-index denoting P1 or P2 is 
suppressed to simplify the notation) is determined by the
expression of $ U_i$ in the adjoint representation.
The matrices $\acal$ and $ \vBB$ must satisfy
\beq
\acal \tau_a \acal^\dagger = \vBB_{ba} \tau_b , \quad
\acal^\dagger \acal = \mati, \quad
\sigma_2 \acal \sigma_2 = \acal^*;
\label{eq:tcons}
\eeq
the first two constraints are required to guarantee 
the invariance of $ \lcal $ under these transformations,
the last constraint follows from the definition of $ \acal$.

Similarly, under reflections
\beq
A_N^a(-y)  = (-1)^{\delta_{N,4}} \tilde\vBB_{ab} A_N^b(y); \qquad
\chi(-y) = -\gamma_5 \bcal \chi (y) ,
\label{eq:refl}
\eeq
with the corresponding constraints
\beq
\bcal \tau_a \bcal^\dagger = \tilde\vBB_{ba} \tau_b , \quad
\bcal^\dagger \bcal = \mati, \quad
\sigma_2 \bcal \sigma_2 = - \bcal^* .
\label{eq:rcons}
\eeq

In addition to the above restrictions the transformations
(\ref{eq:transl},\ref{eq:refl}) must provide a representation
of $ \qcal $. Using the fact
that $ -y = [-(y+L)]+L $ and that $ -(-y) = y$ we find 
\beq
\acal\bcal\acal = \bcal, ~~\vBB\tilde\vBB\vBB = \tilde\vBB; 
\qquad
 \bcal^2=\mati, ~~\tilde\vBB^2 = \mati .
\label{eq:qcons}
\eeq

Finally, under gauge transformations,
$
\chi \to \ocal \chi$, $ \ocal = \exp[ i \omega_a \tau_a ] $
where the $\ocal$ must satisfy
$ \ocal(y+L) = \acal \ocal(y) \acal^\dagger$ ;
$\ocal(-y) = \bcal \ocal(y) \bcal^\dagger $. 

The fermion mass terms may allow for a phenomenologically realistic
low-energy spectrum. The matrix $ \mcal$ is
restricted by requiring invariance under $ \qcal$ and under the local
symmetry group
\beq
[\mcal,\acal]=0 ,~~ \{\mcal,\bcal\}=0; \qquad [\mcal,\tau_a]=0 ; \eeq
also $\mcal=\mcal^\dagger$  and 
$\mcal = -\sigma_2 \mcal^T \sigma_2 $ (form the definition 
in Eq. \ref{eq:defs2}).

The models we consider are then defined by the
Lagrangian $ \lcal$, which specifies the dynamics, as well
as by the matrices $ \vBB, ~ \tilde\vBB, ~ \acal $ and $ \bcal $
that determine the behavior under $ \qcal $.

\section{Light spectrum}

Universal higher-dimensional theories must satisfy the minimum
constraint of generating the experimentally observed light spectrum; because
of this it is of interest to derive the general properties of these
excitations. To this end it proves
convenient to expand the various fields in Fourier modes in the
compact coordinate $y$, the coefficients are then 4-dimensional  fields
for which the action of $ \partial_y$ generates a mass term. It follows 
that all $y$-dependent modes will be heavy (mass $\sim1/L$) and that
light excitations are associated with $y$-independent modes.

The light gauge bosons will be denoted by $A_\mu^{\hat a}$ and the light
fermions by $\chi\up0$, the light modes associated with $A_{N=4}$ behave
as 4-dimensional scalars and will be denoted by $ \phi_{\hat r} =
A^{\hat r}_{N=4} $. Using the $y-$independence of these modes and
the behavior of the field under $ \qcal $ we find 
\bea
A_\mu^{\hat a} &=& \vBB_{\hat a \hat b} A_\mu^{\hat b} = \tilde \vBB_{\hat a \hat b} A_\mu^{\hat b},\cr
\phi^{\hat r} &=& \vBB_{\hat r \hat s} \phi^{\hat s}= -\tilde \vBB_{\hat r \hat s} \phi^{\hat s}, \cr
\chi\up0 &=& \acal\chi\up0 = -\gamma_5 \bcal\chi\up0.
\label{eq:light}
\eea
Light particles are associated with $+1$ eigenvalues of two matrices:
$ +\vBB$ and $ +\tilde\vBB$ for the gauge bosons; 
$ +\vBB$ and $ -\tilde\vBB$ for the scalars; and 
$ \acal$ and $ - \gamma_5 \bcal$ for the fermions.

\section{Simplifying the constraints}

The above set of constraints (\ref{eq:tcons},\ref{eq:rcons},\ref{eq:qcons})
can be simplified by an appropriate choice of
bases. One can then take
\beq
\bcal = \bpm{\mati & 0 \cr 0 & - \mati}\epm,
\quad
\tilde\vBB = \bpm{ \mati & 0 \cr 0 & - \mati}\epm,
\label{eq:newb}
\eeq
(the $0$ and $ \mati$ matrices in $\bcal$ must have the same dimensions;
this is not the case in $ \tilde\vBB$). In this basis we have
\beq
\tilde\vBB \to +1:~\tau = \bpm{\rho&0\cr0&-\rho^*}\epm; \qquad
\tilde\vBB \to -1:~\tau = \bpm{0&\theta\cr\theta^*&0}\epm, 
\label{eq:lbc}
\eeq
It follows from (\ref{eq:light}) that the first of these
expressions determines the couplings between light
fermions and light gauge bosons; similarly the second type of
matrices in (\ref{eq:lbc}) determines the Yukawa couplings
in the light theory.
 
For the fermions, using (\ref{eq:light},\ref{eq:newb}) and 
the constraint $ \chi\up0= - i \sigma\chi\up0{}^c$
we find
\beq
\chi\up0 = \bpm{ \zeta_L \cr -\zeta_L^c }\epm
\eeq

Extracting from $ \lcal$ the terms that contain only light fields, we
find the usual gauge terms for the $ A^{\hat a}$; the gauge-invariant
(under the subgroup associated with the $ A^{\hat a}$) kinetic terms for $ \zeta_L$ 
and $ \phi $, as well as the $ \zeta_L-\phi$ Yukawa terms. The mass
term in $ \lcal$ can generate Dirac and/or Majorana terms for
the $\zeta_L$ depending on the choices of $\acal,~\bcal,~\vBB$ and
$\tilde\vBB$. Note however that the form of $ \lcal$ 
disallows any tree-level potential for
the $ \phi$; it follows that {\em at tree-level} all 4-dimensional 
bosons are either massless or have a mass $\sim1/L$.

If these models are to be phenomenologically
viable, they must be able to generate masses for some of 
the vector bosons at a characteristic scale $v\ll1/L$.
This symmetry breaking step cannot be associated with
the behavior of the fields under $ \qcal $ which necessarily
produces non-zero masses of order $1/L$. But it {\em can}
result from  radiative corrections since these will
generate a non-vanishing (effective) potential for the $\phi$
at $ \ge1$ loops. This opens the possibility that these models will undergo
two stages of symmetry breaking: the first generated by the behavior 
under $ \qcal $ and the second, at a presumably lower scale, generated
radiatively by the scalars. Though we have not yet succeeded in generating
a phenomenologically viable theory along these lines, we do have
examples where these features are realized. 

\section{$\ui$ example}

We look for a $\ui$ gauge theory~\cite{Grzadkowski:2005rz}
where the $\qcal$ transformations induce the
breaking $\ui\to$nothing while generating a massless (at tree-level) scalar.
We include first a single fermion flavor, then the constraints
are all satisfied by the choices $\vBB = -\tilde \vBB= 1$ and
\beq
\bcal = \sigma_3, \quad
\acal = \bpm{ \cos u & - \sin u \cr \sin u & \cos u }\epm ,\quad 
\mcal = \bar\mu \sigma_2,\quad
\tau = g \sigma_2,
\eeq
where $ \bar\mu$ denotes a mass parameter and $g$
the gauge coupling.

\begin{figure}[ht]
\centering
\includegraphics[width=3in]{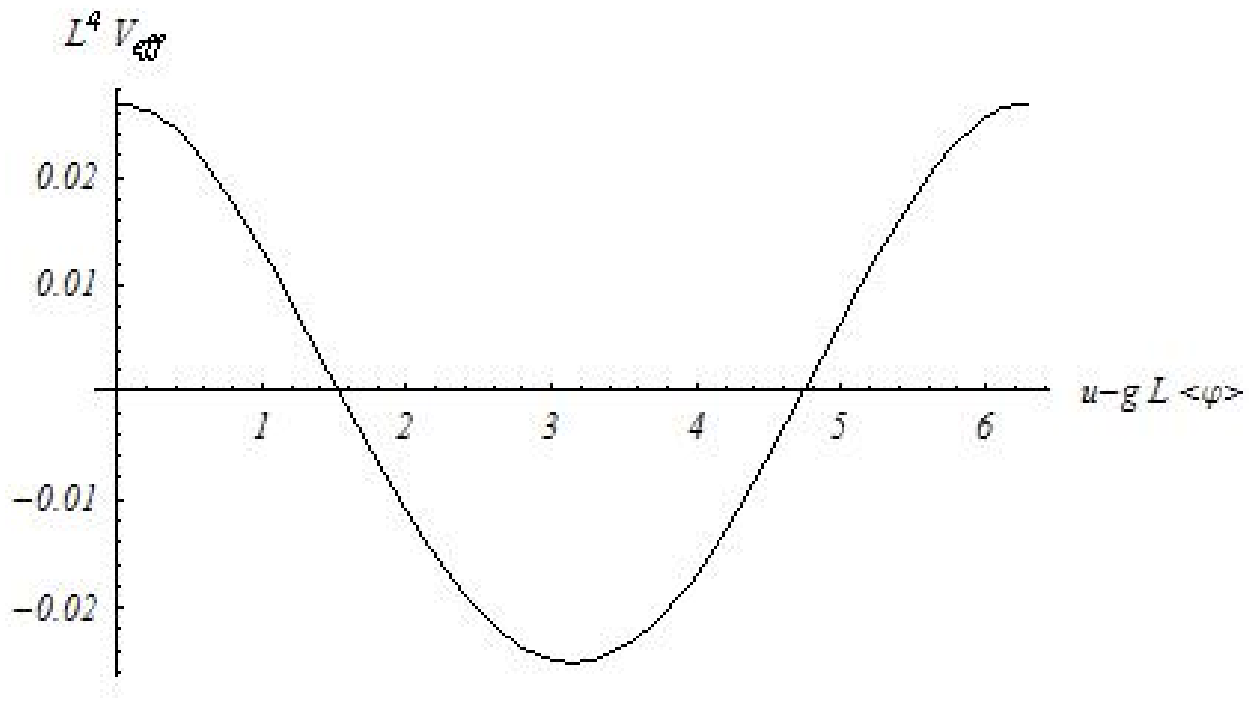}
\includegraphics[width=3in]{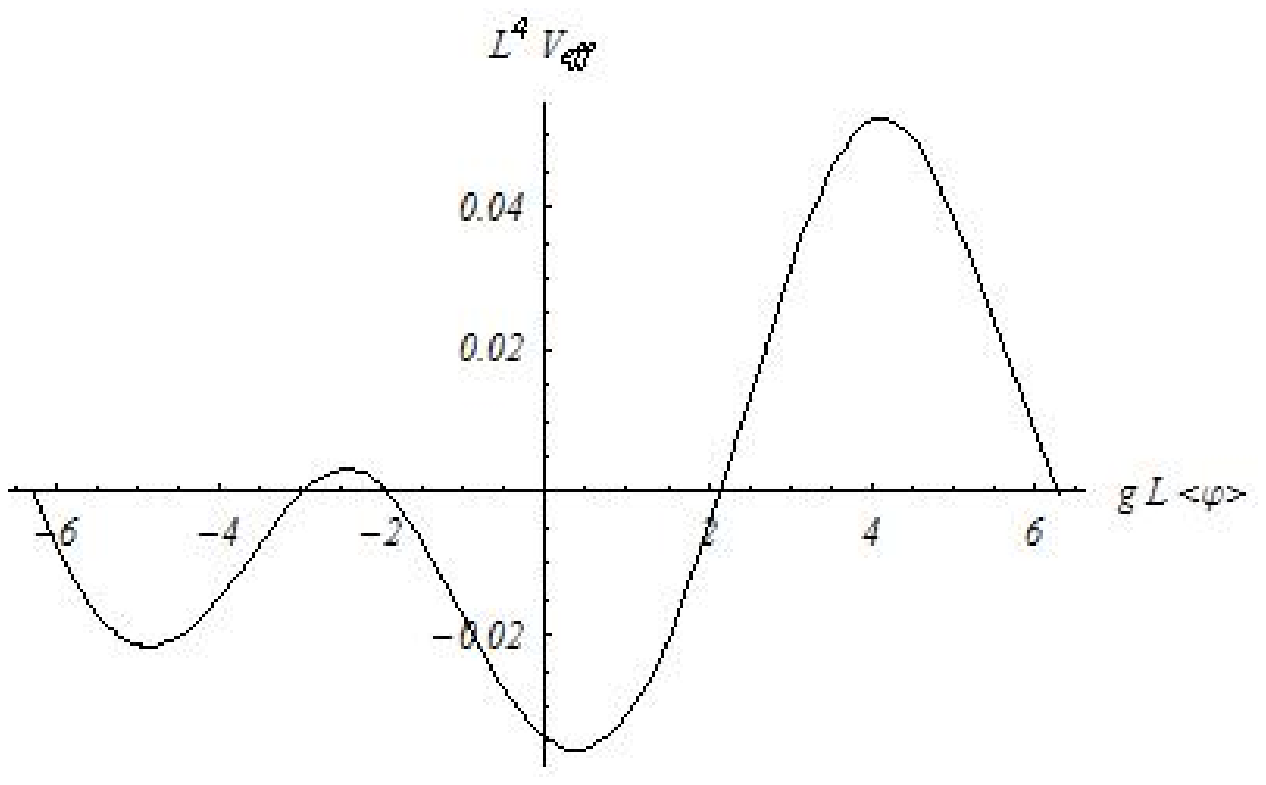}
\caption{The effective potential for the $\ui$
model. Left: single fermion species
with $ \bar\mu L = 0.01$. Right:
two fermion species with
$\bar\mu_1 L = 0.01,~ \bar\mu_2 L = 0.008$,
$u_1 =2.3,~u_2=0$ and charges $1$ and $-1/2$;
note that for $u_1$ the tree-level mass is $\simeq 2.3/L$.}
\label{fig:f1}
\end{figure}

The tree-level light spectrum of this model consists
of a Majorana fermion with mass $ \sqrt{\bar\mu^2+(u/L)^2}$
and a neutral massless scalar. The 1-loop effective potential
is given by~\cite{Hatanaka:1999sx}
\beq
V_{\rm eff} = \inv{4\pi^2 L^4} \Re\left[
{\rm Li}_5(\zeta)
+ 3 x {\rm Li}_4(\zeta)
+ x^2 {\rm Li}_3(\zeta) \right],
\label{eq:u1veff}
\eeq
where $ x= \bar\mu L $ and $ \zeta= \exp\left[ 
- x + i \left( u - g L \vevof\phi \right)\right]
$
which is plotted for one and two fermion
species in Fig. \ref{fig:f1}. In the two-flavor case the
presence of a heavy fermion can lead to a
\vev\ $ \vevof\phi \ll 1/L$.

\section{$\su2$ example}

We look for an $\su2$ theory where the light sector is invariant under a $\ui$
subgroup and contains one complex scalar. We include
2 fermion doublets.

All the constraints are satisfied by the choices
\bea
&& \begin{array}{ll}
\vBB = \mati_3,
&
\tilde\vBB = \dm{-1,+1,-1},
\cr&\cr
\bcal = \dm{\mati_4,-\mati_4},
&
\acal = \dm{\mati_2,-\mati_2,\mati_2,-\mati_2} ,
\cr&\cr
\mcal = \bpm{0&\mubf\cr\mubf&0}\epm,
&
\mubf =\frac i2\bpm{\bar\mu_+\sigma_1&0\cr0,
&\bar\mu_- \sigma_1}\epm
\end{array} \cr
&&\cr
&&\cr
&&
\tau_1 = \frac i2\bpm{0&-\rho\cr\rho&0}\epm,
~~
\tau_2=\frac 12\bpm{\rho&0\cr0&-\rho}\epm,
~~
\tau_3=\frac 12\bpm{0&\mati_4\cr\mati_4&0}\epm,
\eea
where 
$\rho = \dm{-1,+1,-1,+1}$ and  $\bar\mu_\pm$ are real.

Using (\ref{eq:light}) these expressions show that the
light (tree-level spectrum) consists of a $\ui$ gauge boson,
one Dirac fermion of mass $ \bar \mu_+$ and one charged scalar.
The one-loop effective potential has a form similar to 
(\ref{eq:u1veff}) and is plotted in fig. \ref{fig:f2}.
This plot seems to indicate that the $\ui$ symmetry is broken and that
there are in fact no massless vector bosons. This
is not the case: the masses of the vector Fourier modes are 
$m_n = 2\pi n/L+2 g \vevof\phi$; at tree level $\vevof\phi=0$ so
$ m_n\up{\rm tree} =  2\pi n/L$ and we identify the $\ui$ gauge boson with
the $n=0$ mode. At one loop $ \vevof\phi = \pm \pi/(L g) $ so that 
$ m_n\up{1\rm loop} =  2\pi (n \pm1)/L$ and we identify the $\ui$ gauge boson with
the $n=\mp1$ mode.

\begin{figure}[ht]
\centering
\includegraphics[width=4in]{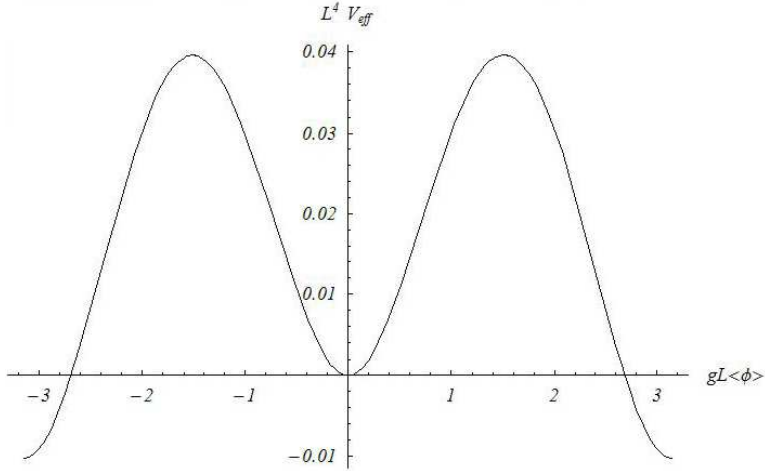}
\caption{The effective potential for the $\su2$
model; $\bar\mu_- L = 2.0,~ \bar\mu_+ L = 0.001$.}
\end{figure}
\label{fig:f2}

These results suggest that with the proposed transformation properties
(\ref{eq:transl}, \ref{eq:refl})
these theories could generate the correct
low-energy physics. The difficulty lies in constructing an effective potential
that has the right value of $\vevof\phi$. This apparently necessitates the introduction of
additional fermion representations which, however, need not be 
light and  would not spoil the light spectrum. These
models are currently under investigation.

\end{document}